%% file: bbbar_herab.tex
\documentclass{elsart}
\usepackage{epsfig}
\input{macros.tex}

\newcommand{\Figdir}{}
\newcommand{\myfiglabel}[1]{\label{#1}}
\journal{European Journal of Physics C}

\begin{document}

\begin{frontmatter}
\vbox{\normalsize
\noindent
\rightline{\hfill {DESY-02-076 }}\\
\rightline{\hfill {hep-ex/0205106}}
\vskip 3pt
\rightline{\hfill {31$^{st}$-May-2002}}
}

\title{ Measurement of the {\boldmath \bbar } Production Cross 
Section in 920 GeV Fixed-Target Proton-Nucleus Collisions
}

\author{The \hb Collaboration}

\input authors.tex


\begin{abstract}
Using the \hb detector, the \bbar\ production cross section has been
measured in 920 GeV proton collisions on carbon and titanium targets. The \bbar\ production was 
tagged via inclusive bottom quark decays into \jpsi\, by exploiting the longitudinal separation of \jpsill\ decay 
vertices from the primary proton-nucleus interaction. Both \ee\ and \mm\ channels have been reconstructed 
and the combined analysis yields the cross section 
$\sigbbar\ = 32 ^{+14} _{-12} \mathrm{(stat)} \, ^{+6} _{-7} \mathrm{(sys)} \ \mathrm{nb/nucleon}$.
\end{abstract}

\begin{keyword}
hadroproduction, \ bottom mesons
\PACS 13.85.Ni \ 13.85.Qk \ 13.20.He \ 24.85.+p
\end{keyword}

\end{frontmatter}

\clearpage

\section{Introduction}

The theoretical description of heavy quark hadroproduction has been a 
subject of great attention in recent 
years~\cite{nason2001,nason1998,vogt2001,frixio2001}. For fixed target 
experiments~\cite{nason1998,vogt2001}, the predictions, based on perturbative 
QCD, are compatible with experimental results from pion and proton beams, 
but both theoretical and experimental uncertainties are large. Only two 
measurements of the \bbar\ production cross section in proton-nucleus 
$p+A$ interactions were previously reported~\cite{E789bb,E771bb}.
Moreover, in collider experiments, the comparison between data and QCD 
predictions is not satisfactory~\cite{frixio2001}. 

The \hb experiment is designed to identify $B$-meson decays 
in a dense hadronic environment, with a large geometrical coverage. 
Interactions are produced on target wires in the halo of the 920 GeV HERA 
proton beam. The
\bbar\ production cross section (\sigB) on a nucleus of atomic number $A$
is obtained from the inclusive reaction
\begin{equation}
pA \rightarrow \bbar \; X  \ \ \textrm{with} \ \ \bbar \rightarrow 
\jpsi \, Y \rightarrow  (e^+ e^- / \mu^+ \mu^-) Y .
\end{equation}
The \B-hadron decays into \jpsi\ (``\bjpsi " in the following) are
distinguished from the large prompt \jpsi\ background by exploiting the \B\
lifetime in a detached vertex analysis. 
We select \bjpsill\ decays in both the muon and electron
channels and perform a combined \bbar\ production cross section
measurement.

In order to minimize the systematic errors related to detector and trigger
efficiencies and to remove the dependence on the absolute luminosity
determination, the measurement is performed relative to the
known prompt \jpsi\ production cross section \sigP\ ~\cite{E789dj,E771dj}. Our 
measurement covers the \jpsi\ Feynman-$x$ (\xf ) 
range $-0.25  \le \xf \le  0.15$. Within our acceptance, the \B\ to prompt cross section ratio 
can be expressed as:
\begin{equation}
\frac{\Delta \sigB}{\Delta \sigP}  = \frac{N_B}{N_P} \, {1 \over {  \effR \, 
  \effBz \, \Br{\bbjX}}} \ ,
\label{eq:sigbbar}
\end{equation}
where $\Delta \sigB$ and $\Delta \sigP$ are the \bjpsi\
and prompt \jpsi\ cross sections limited to the mentioned \xf\ range, 
$N_B$ and $N_P$ are the observed number of detached \bjpsi\ and prompt \jpsi\ decays. 
\effR\ is the relative detection efficiency of \bjpsi\ with respect to 
prompt \jpsi , including contributions from the trigger, the dilepton 
vertex and the \jpsi\ reconstruction. 
\effBz\ is the efficiency 
of the detached vertex selection. The branching ratio \Br{\bbjX}\ in hadroproduction is assumed to be the 
same as that measured in $Z$ decays, with the value $2 \cdot (1.16\pm 0.10)\%$~\cite{PDG}. 

The prompt \jpsi\ production cross section per nucleon, $\sigma(pN\to \jpsi X)
= {\sigP}/{A^\alpha}$, was previously measured by two fixed target
experiments~\cite{E789dj,E771dj}. After correcting for the most recent
measurement of the atomic number dependence ($\alpha = 0.955 \pm 0.005$
~\cite{E866}) and rescaling~\cite{E771dj2} to the \hb c.m.s. energy,
$\sqrt{s}$ = 41.6 GeV, we obtain a reference prompt \jpsi\ cross section of
$\sigma(pN\to \jpsi X) = (357 \pm 8 \mathrm{(stat)} \pm 27 \mathrm{(sys)})\
\mathrm{nb/nucleon}$. About 70\%~\cite{E789dj} of the \jpsi\ are produced in
the kinematic range covered by our measurement. 

Since no nuclear suppression has been observed in $D$-meson
production~\cite{Ddep} and a similar behavior is expected 
in the \B\ channel~\cite{Adep}, 
we assume $\alpha = 1.0$ for the \bbar\ production cross section 
results presented here; i.e., $\sigB\ =\sigma(pN\to \bbar)\cdot A$.

\section{Detector, Trigger and Data sample}
\label{sec:det}

A side view of the \hb \cite{hb_tdr,hb_oct} spectrometer is shown in
Fig.~\ref{fig:detect}. The spectrometer has a large geometrical coverage,
from 15 mrad to 220 mrad in
the bending plane and from 15 mrad to 160 mrad in the vertical plane. 
\begin{figure*}
\resizebox{0.95\textwidth}{!}{%
  \includegraphics{\Figdir 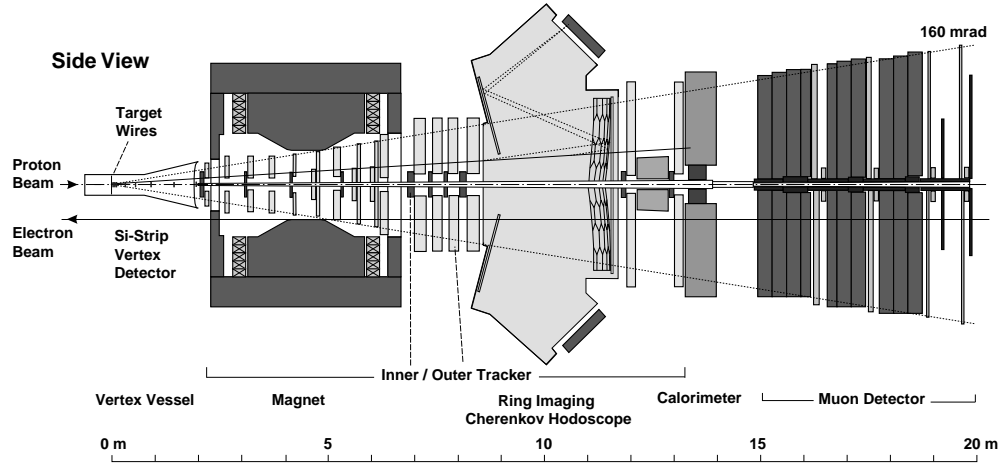}
}
\caption{Side view of the \hb detector. }
\myfiglabel{fig:detect}       
\end{figure*}

The target assembly~\cite{hb_targ} consists of two wire stations separated
by 4 cm along the beam line, each containing 4 target wires of
different materials. A servo system automatically steers the target wires
during a run in order to maintain a constant interaction rate. 
The Vertex Detector System (VDS)~\cite{hb_vdet} is realized by a system of 20 
Roman pots containing seven planar stations (4 stereo views) of double-sided silicon 
micro-strip detectors (50 mm $\times$ 70 mm sensitive area, 50 $\mu$m pitch) that are operated 
in a vacuum vessel at 10 to 15 mm distance from the proton beam. An additional 
station is mounted immediately behind the 3 mm thick Al window of the vacuum 
vessel.

A dipole magnet with 2.13 Tm field-integral houses a first set of tracking 
stations, followed by a second set extending to 13 m downstream of the
interaction region. 
To cope with the large particle flux
gradient radial to the beam, the tracker is divided into 
a fine grained Inner Tracker (ITR)~\cite{hb_msgc} and a
large area Outer Tracker (OTR)~\cite{hb_tdr,hb_otr}. 
The ITR uses micro-strip gas chambers (typical pitch of 300 $\mu$m) 
with gas electron multipliers. The OTR
uses honeycomb drift cells with wire pitches of 5 mm near the beam and
10 mm away from the beam.

The particle identification is performed by a ring imaging Cherenkov hodoscope
(RICH)~\cite{hb_rich}, an electromagnetic calori\-meter (ECAL)~\cite{hb_ecal}
and a muon detector (MUON)~\cite{hb_mu}. The RICH detector uses $\mathrm{C_4F_{10}}$ 
as radiator. The focal planes of the detector (above and below the beam line, 
respectively) are read out by multianode photomultipliers.
The ECAL is based on ``shashlik" sampling calorimeter technology,
consisting of scintillator layers sandwiched between metal absorbers. 
In the radially innermost section, W is used in the absorber, and Pb everywhere else.
The MUON system consists of 4 tracking stations located in the most downstream portion
of the detector, at different depths in iron and iron-loaded absorbers. It is built
from gas-pixel chambers in the radially innermost region and from proportional
tube chambers, some with segmented cathodes (pads), everywhere else.

The data sample presented in this analysis was acquired in a short physics run
during the \hb commissioning period in summer 2000, at $\approx$5 MHz interaction
rate, with a maximum of two target wires operated simultaneously and separated
by  4 cm along the beam direction. The two wires were made of carbon
(1000 $\mu$m longitudinally and 100 $\mu$m transversely) and titanium (500
$\mu$m and 50 $\mu$m, respectively). 

The data were collected by triggering on dimuon and dielectron signatures. The MUON
pretrigger candidates were based on a double pad chamber
coincidence~\cite{hb_mu_pre},  while the ECAL pretrigger candidates were
defined by ECAL clusters with a transverse energy $E_T > 1.0$
GeV~\cite{hb_ecal_pre}. The First Level Trigger (FLT) required two pretrigger
candidates of the same type and forwarded these to the Second Level Trigger
(SLT). The SLT is a software filter~\cite{hb_slt} running on a farm of 240
PCs. Starting from the pad coincidences and high-$E_T$ ECAL
clusters, a fast hit-counting algorithm and a
simplified Kalman filter were applied to the OTR and VDS data to confirm the
lepton pair candidates.  An invariant mass cut of $M > 2.0$~\mgev\ and
an unlike-sign track requirement were also applied
in the electron channel.  The data from accepted events were assembled and
sent to the online reconstruction farm~\cite{hb_4lt}, 
consisting of 100 dual-CPU PCs.
The whole trigger chain allowed a reduction of the initial 
interaction rate of 5 MHz to a final output rate of 20 Hz~\cite{hb_daq}.
A total of $\approx$450,000 dimuon and $\approx$900,000 dielectron 
candidates were recorded under these conditions. 

At the time of data taking, the ITR and the MUON pixel chambers were not 
included in the trigger. As a consequence, the forward hemisphere of 
the proton-nucleus c.m.s. is reduced in this measurement, compared to the 
full \hb acceptance.

\section{Monte Carlo Simulation}
\label{sect:mc}

A Monte Carlo (MC) simulation is used to determine the efficiency terms
in Eq.~\ref{eq:sigbbar} and to estimate the prompt
\jpsi\ background contribution to the \bjpsi\ decay channel. 

The simulation of heavy quark ($Q$) production is achi\-eved, first, 
by generating the basic process $pN \to Q \bar Q X$ including hadronization, 
using the {\sc Pythia 5.7} event generator \cite{pythia}; secondly, the remaining  
part of the process ($X$) is given as an input to the {\sc Fritiof 7.02} package~\cite{fritiof} 
to simulate further interactions inside the  nucleus.

To describe the prompt \jpsi\ kinematics accurately, 
the generated events 
are weighted according to the known prompt \jpsi\ differential 
cross sections 
($d\sigma /d\pt^2$ and $d\sigma / d \xf$) 
measured in proton-gold collisions~\cite{E789dj}. These 
results were obtained in the positive \xf\ region, while our measurement 
covers the range $-0.25 \le \xf \le 0.15$. MC studies based on the Color 
Octet Model~\cite{com} of charmonium production show a symmetric \xf\ distribution 
of prompt \jpsi\ decays. We therefore use the experimental parameterization~\cite{E789dj} to extrapolate 
to the full \xf\ space. The model dependence of the generated \pt\ spectrum is 
of less relevance since our \pt\ acceptance is essentially flat. 

For the \bbar\ MC simulations, the events generated by {\sc Pythia} are
weighted  according to a model with various contributions. First,  the
generated \B\ quark kinematics  (\xf\ and \pt ) are given by the computation
of M. Mangano {\it et al.} \cite{mang} using  the most recent
next-to-next-to-leading-logarithm (NNLL)  MRST parton distribution
functions~\cite{mrst} with a \B\ quark mass of  $m_{\B} = 4.75$~\mgev\ and a
QCD renormalization scale  $\mu=\sqrt{m^2_\B + \pt^2}$. Second, the intrinsic
transverse momenta  of the colliding quarks are smeared with a Gaussian
distribution  leading to $\langle k^2 _T \rangle = 0.5 \ \egev^2/c^2$
\cite{chay}.  Finally, the \B\ fragmentation is described by a Peterson
function  \cite{peterson} with a parameter $\epsilon = 0.006$ \cite{E789bb}.
The subsequent \B-hadron production and  decay are controlled by the {\sc
Pythia }  default parameters.  The \B-hadron average lifetime is taken from
Ref.~\cite{PDG}:  $\tau_\B = 1.564 \pm 0.014 \, \textrm{ps}$.

The sensitivity of the final result on the \bbar\ cross section within our
acceptance (\dsigbbar ) has been determined by varying
the following \bbar\ MC model parameters: 
the parton distribution functions (from MRST to CTEQ5 \cite{cteq5}),
the \B\ quark mass (in the range
$m_\B \in [4.5, 5.0]$~\mgev), the QCD renormalization scale (from 
$0.5\sqrt{m^2_\B + \pt^2}$ to $2\sqrt{m^2_\B + \pt^2}$), the fragmentation
function (from the Peterson form~\cite{E789bb,epsil,aleph} with parameter
$\epsilon \in [0.002, 0.008]$, to the Kartvelishvili form \cite{kart}
with parameter $\alpha_\beta = 13.7 \pm 1.3$ \cite{aleph}), the 
intrinsic transverse momentum distribution 
(with $\langle k^2 _T \rangle$ in the range [0.125,2.0] $\egev^2/c^2$) and
the fraction of \B-baryons produced in the \B\  hadronization process in the
range [0, 12]\%. 
The observed variations in the detection efficiencies have been included in the
systematic error.

The generated particles are propagated through the geometry and material 
description of the detector using  the {\sc Geant 3.21} package 
\cite{geant}. A simulation of the detector response to particles is 
achieved by reproducing the digitization of electronic signals, with 
a realistic description of hit efficiencies and problematic channels. 
The MC events are subjected to a full trigger simulation 
and reconstructed with the same algorithms as the data. 

\section{{\boldmath \jpsi} Event Selection}
\label{sec:sel}

Since the observed number of prompt \jpsi\ decays is used as a normalization
 factor of the \bbar\ cross section measurement,
 we begin by selecting and counting the number of \jpsi\ decays ($N_P$), 
before applying the detached vertex analysis. 
The lepton reconstruction in the OTR is seeded with the
dilepton trigger track candidates; moreover a matching criteria is applied between 
the reconstructed track and the trigger track candidate in
both the OTR and the VDS.
The \jpsill\
selection and counting procedure differs between the muon and
electron channels, due to differences in the background levels, shapes and
triggering conditions.

\subsection{ \boldmath \jpsimm\ }

Three criteria are used to select \jpsimm\ decays and to purify the
reconstructed sample from non-\jpsi\ background: a dimuon vertex requirement
and muon identification cuts in both the MUON and the RICH systems. The cuts
are chosen to give the best signal significance
($S/\sqrt{B}$) on the number of seen \jpsi\ ($S$) with the observed background ($B$). 
The resulting spectrum is shown in Fig.~\ref{fig:fullmuon},
with $N_P = 2880\pm 60$ prompt \jpsimm\ decays. The like-sign spectrum shown
in Fig.~\ref{fig:fullmuon} is obtained from the same set of triggered
events: the small discrepancy in number of reconstructed events in the
background regions arises from the difference in trigger acceptance between
the two cases and from physics contributions to the unlike-sign spectrum
(Drell-Yan, open charm production). 

%
%
%
%
\begin{figure}
\resizebox{0.75\textwidth}{!}{%
\includegraphics{\Figdir 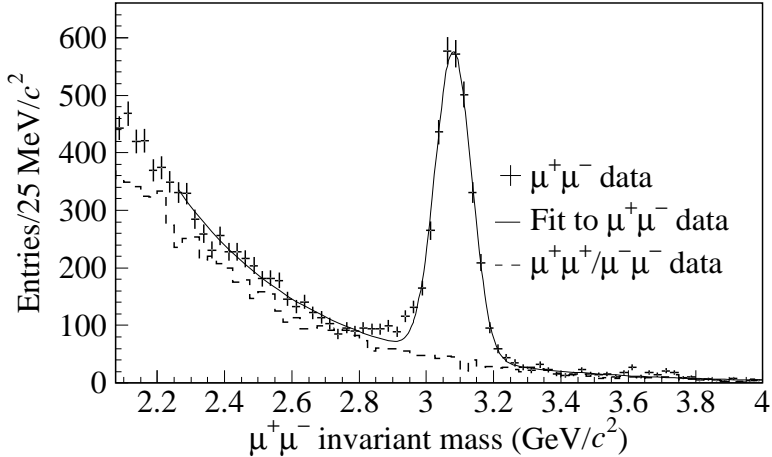}
}
\caption{The \mm\ invariant mass spectrum, after the \jpsi\ selection cuts. The fit (solid line) assumes a Gaussian signal and an exponential background. The like-sign spectrum (dashed line) shows small discrepancies in the background regions (see text); it is not used in the analysis and serves only for illustration purposes. }
\myfiglabel{fig:fullmuon}
\end{figure}
%
\subsection{ \boldmath \jpsiee\ }

The selection of \jpsiee\ decays is more complex due to very large
background contributions, mainly from pions interacting in the ECAL
and hadrons overlapping with energetic neutral showers.  Due to such
background, a clear \jpsi\ signal can be reconstructed only by means of 
strong electron identification requirements. Electron identification in 
the ECAL is based both on the ratio of the cluster energy $E$  to the 
momentum $p$ from  tracking ($E/p$) and on the search for electron 
bremsstrahlung signals:

- the $E/p$ distribution is established for a purified \jpsiee\ sample by
using a  double-bremsstrahlung requirement as described in the following
paragraph. The $E/p$ spectrum  is compatible with a Gaussian distribution of
mean $1.00$ and width $\sigma\approx$9\%;

- bremsstrahlung photons emitted upstream of the magnet maintain the
original  electron direction; thus they can be used to correct the electron
momentum at the vertex  and they also provide a clean electron signature
(bremsstrahlung tag).

The \ee\ invariant mass distribution is shown in Fig.~\ref{fig:eejpsi}(a),
requiring only that $E/p$ be within 1 $\sigma$ from unity for each
track. Figs~\ref{fig:eejpsi}(b,c) show the improvements in signal significance
that are obtained when the bremsstrahlung selection 
is added to the $E/p$ requirement.

%
%
%
%
\begin{figure}
\resizebox{0.75\textwidth}{!}{%
\includegraphics{\Figdir 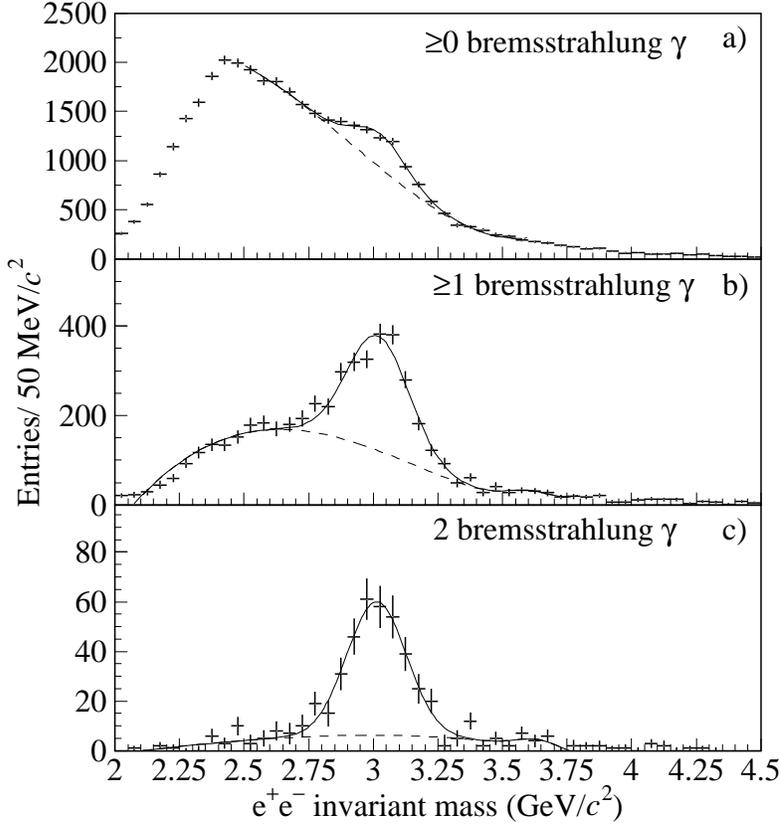}
}
\caption{The \ee\ invariant mass spectra, with an $E/p$ cut at 1 $\sigma$ on both $e^+$ and $e^-$ tracks; 
(a) no bremsstrahlung requirement; (b) with at least 1 identified bremsstrahlung photon; (c) with two identified 
bremsstrahlung photons. The fits (solid lines) assume a Gaussian signal and a polynomial shape for the background. 
The background shapes (dashed lines) differ from the dimuon case (Fig.~\ref{fig:fullmuon}), due to $E_T$ and invariant mass cuts in the trigger.}
\myfiglabel{fig:eejpsi}
\end{figure}
%
\begin{table*}
\begin{center}
\begin{scriptsize}
\begin{tabular}{|c|c|c|c|c|}
\hline
 Bremsstrahlung     & \multicolumn{4}{c@{\ \ \ }|}{$E/p$ cut in units of $\sigma$} \\ 
 requirement  & $0.5 \sigma $ & $1 \sigma$ & $2 \sigma$ & $3 \sigma$ \\ \hline
 None &$895 \pm 75$&$2553 \pm 184$ & $5362 \pm 292$& [$5710 \pm 380 \pm 280$]\\
$ \ge 1 $&$519 \pm 45 \pm 11$&$1420 \pm 70 \pm 48$ & $2851 \pm 108 \pm 48$&$3304 \pm 148 \pm 188$\\
$ 2  $&$106 \pm 13 \pm 2$&$ 308 \pm 24 \pm 6$ & $ 587 \pm 38 \pm 14$&$ 661
\pm 40 \pm 29$ \\ \hline
$\epsilon_{\mathrm{brems}}$&$0.34\pm0.03 \pm 0.01$&$0.35\pm0.02 \pm 0.01$ & $0.33\pm0.02 \pm 0.01$&$0.33\pm0.02 \pm 0.02$ \\  
\hline 
\end{tabular}
\end{scriptsize}
\caption{\label{bbarelectr:JPSIcount}\small The number of prompt \jpsiee\ with 
different bremsstrahlung requirements and different $E/p$ cuts 
on both tracks. The bremsstrahlung tag probability for a single electron
is reported for each $E/p$ cut. The value at $3
\sigma$ in square brackets is obtained by extrapolation from samples 
with stronger electron identification cuts (see text).}
\end{center}
\end{table*}

Table~\ref{bbarelectr:JPSIcount} lists the number of prompt \jpsi\ signal
events found for different $E/p$ and  bremsstrahlung requirements. Using these
\jpsi\ sets one can measure the bremsstrahlung tag probability
for a single electron redundantly,
resulting in an average $\epsilon_{\mathrm{brems}} = 0.34 \pm 0.02\mathrm{(stat)}
\pm\,0.02 \mathrm{(sys)}$. This measured tag probability is in good agreement with
expectations from MC simulations. The systematic uncertainties include the
observed fluctuations when varying the fitting functions and range used to
estimate the amount of \jpsi\ events in the invariant mass spectra.

The good knowledge of the particle identification efficiencies allows 
us to infer the number of prompt \jpsi\  present in a sample where 
looser identification cuts have been applied and where no clear \jpsi\ signal 
is directly visible. Under such conditions, the normalization factor 
$N_P$ can be obtained while preserving reasonable statistics for the 
final detached vertex analysis, which relies only on the vertex separation 
cuts for the background rejection, as will be shown in Sect.~\ref{sec:anal}. 

The total number of prompt \jpsi\  in our sample with no bremsstrahlung tag
requirement and with a 3$\sigma$ $E/p$ cut
is $N_P = 5710 \pm 380 \mathrm{(stat)} \pm 280 \mathrm{(sys)}$ 
(entry in square brackets in Table~\ref{bbarelectr:JPSIcount}).

\section{Detached Vertex Analysis}
\label{sec:anal}

The long decay length of \B-hadrons is used to separate the \bjpsi\ events
from the  prompt \jpsi\ and to further reduce the non-\jpsi\ background. The
decay length ($\Delta z$),  defined as the distance along the beam axis
between  the \jpsi\ decay vertex and the closest wire (primary production
point),  is shown in Fig.~\ref{fig:dz} for a purified sample of \jpsiee\
events (2 bremsstrahlung requirement).  The width of the prompt \jpsi\ vertex
distribution is more than 10 standard deviations  smaller than the mean decay
length of triggered \B-hadrons ($\approx$0.8 cm).  Given the achieved vertex
resolution, a detached vertex cut proves to be efficient in the signal
selection. Additionally, a cut on the minimum impact parameter of both leptons
to the production vertex ($I_v$) or to the wire ($I_w$) is applied in the
detached  vertex selection. The minimum impact parameter distribution of 
prompt \jpsiee\ decay leptons (purified sample) to the wire is compared to the corresponding  distribution from \bjpsi\ decay leptons (dashed line) in Fig.~\ref{fig:eeimpact}, illustrating the potential gain in \bjpsi\ signal-purity when applying a minimum impact parameter cut. 

%
%
%
%
\begin{figure}[t]
\resizebox{0.75\textwidth}{!}{%
\includegraphics{\Figdir 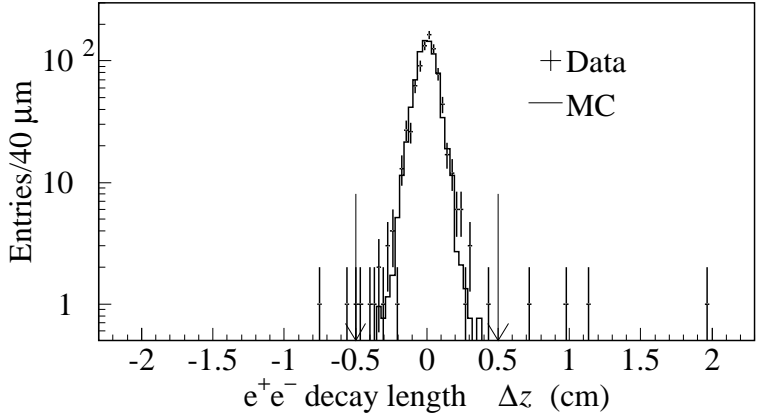}
}
    \caption{ \small The decay length ($\Delta z$) of purified 
$\jpsi \to \ee $ events (2 bremsstrahlung sample, $\approx 15 \%$ background) 
for real data and compared to MC prompt \jpsi\ events. The resolutions of both data and MC distributions are in agreement (715~$\pm$~24~$\mu$m and 721~$\pm$~10~$\mu$m, respectively). The arrows mark the cuts applied in the detached vertex analysis.}  
\myfiglabel{fig:dz}
\end{figure}

The prompt \jpsi\ which survive the detached vertex cuts cannot be
distinguished from  \bjpsi\ events and their contribution to the detached \B\
signal has to be determined from MC. A study has been performed to ensure that
the simulation tool reliably reproduces the real data for the physical
quantities defining the detached selection cuts, as illustrated by the good
MC-data agreement in Fig.~\ref{fig:dz} and Fig.~\ref{fig:eeimpact}. In the
simulation of the \mm\ channel, a Gaussian distribution has been added to the
standard MC track-slopes, increasing the slope errors by 20\% in average, in
order to  match the parameters observed in data and resulting in a similar
good MC-data agreement as  in the \ee\ case.

%
%
%
\begin{figure}
\resizebox{0.75\textwidth}{!}{%
\includegraphics{\Figdir 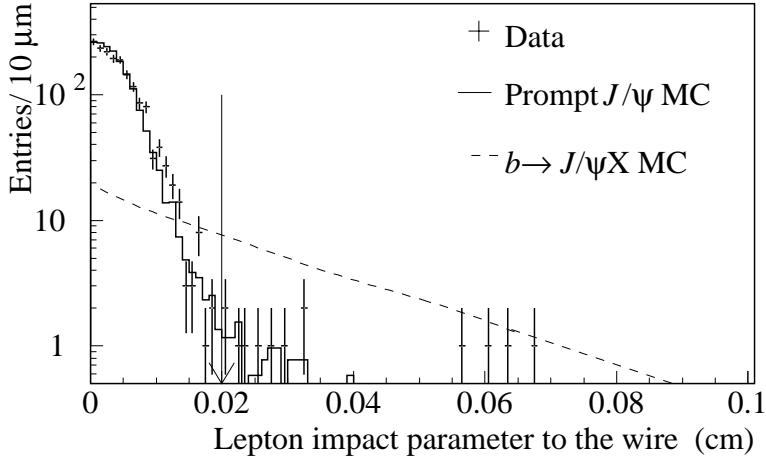}
}
\caption{The minimal impact to the wire ($I_w$) of the two lepton
tracks for purified $\jpsi \to \ee $ events (2 bremsstrahlung sample) data 
and in MC for \jpsiee\ events. 
The expected shape from \bjpsi\ decays is shown in dashed line (arbitrary scale). The arrow marks the cut applied in the \ee\ detached event selection.  
}
\myfiglabel{fig:eeimpact}
\end{figure}

The optimization of the detached vertex cuts is achieved by maximizing the
ratio $S/\sqrt{B}$, where $S$ is the number of 
accepted signal events in the MC \bjpsi\  sample and 
$B$ the number of background events in real data, 
observed in the whole upstream region and downstream in the side bins of
the \jpsi\  invariant mass signal.
The systematic errors in the final \sigbbar\ result take into account the
variations in the signal estimated under different sets of cuts found by the
optimization procedure. 

The main background contributions expected in the detached sample are
due to combinatorics (bad vertex or track reconstruction) and
to double semileptonic $c\bar c$ or \bbar\ events.
The combinatorial yield is estimated through the observed events in 
the region upstream of the primary interaction 
(unphysical region), while the charm background
level is estimated by means of MC simulations,
assuming a $c\bar c$ production cross section of 
$40 \ \mu$b/nucleon for our 920 GeV proton beam energy~\cite{frixio}. 
The \bbar\ background yield is estimated by studying the MC mass spectrum 
of \bbar\ events surviving trigger and selection cuts, with a contribution relative 
to the observed yield of \bjpsill\ events.   

\subsection{ \boldmath \bjpsimm\ }

The detached vertex cuts found by the optimization procedure in the dimuon channel are: a minimum 
decay length of 7.5 times the uncertainty on the secondary vertex position~\footnote{The typical cutoff is at about 5 mm, 
with some events down to 3 mm being accepted.}, a minimum track impact parameter 
to the assigned primary vertex of 160 $\mu$m and a minimum track impact parameter to the assigned 
wire of 45 $\mu$m. 

Only 11 events survive these cuts downstream of the primary interaction region with invariant 
mass above 2.1 \mgev\ (see Fig.~\ref{fig:mmdetached}). Only a single event is 
found upstream of the primary interaction. 
%
%
%
%
\begin{figure}
\resizebox{0.75\textwidth}{!}{%
\includegraphics{\Figdir 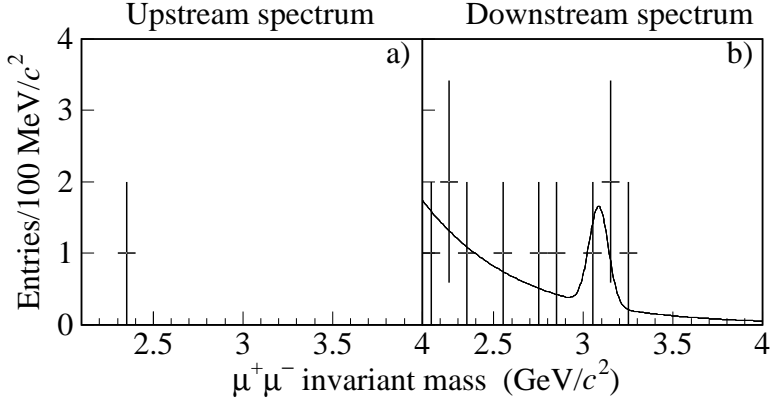}
}
\caption{
The upstream (a) and downstream (b) invariant \mm\ mass spectrum 
after detached event selection. 
The downstream curve shows the result of the unbinned likelihood fit, in which
the yields of background and signal contributions and the background slope, 
as defined in the text, were left as free parameters.  
}
\myfiglabel{fig:mmdetached}
\end{figure}

In order to use the full information of the detached invariant mass spectrum, 
an
unbinned likelihood fit is performed, using the Gaussian parameters of the
prompt \jpsi\ signal together with an exponential background contribution with 
free slope. The output of the fit shown in Fig.~\ref{fig:mmdetached} yields 
1.9$^{+2.2}_{-1.5}$ \bjpsi\ events. 
The background slope obtained from the fit is compatible with the
simulated charm and bottom quark background shape, although the statistics are low.

From simulation, the expected background from prompt \jpsi\ decays is
negligible.  The estimated background contributions of semileptonic charm and
bottom quark decays,  together with the single event expected from combinatorial
background (seen upstream) and the number of fitted signal events, are
compatible with the 11 events observed downstream  of the primary interaction
region.

To determine \dsigbbar\ in our \xf\ range, the prompt \jpsi\ and \bjpsi\ MC
events are submitted to the same analysis chain used for real data. From
simulation we obtain the efficiency terms entering in the cross section
formula:  $\effR \cdot \effBz = 0.41\pm 0.01$. The corresponding \bbar\ cross
section  measured in the \mm\ channel is  $\dsigbbar =\sigB /A =
16^{+18}_{-12}$ nb/nucleon,  obtained by using the weighted average of our
target materials. All parameters  contributing to the measurement
(Eq.~\ref{eq:sigbbar}) are summarized in  Table~\ref{tab:comp}.

\subsection{ \boldmath \bjpsiee\ }
The cut optimization procedure in the \ee\ channel results in the following
criteria: a minimum decay length of 0.5 cm, a  minimum track impact parameter 
to the assigned wire of 200 $\mu$m or alternatively an isolation of the lepton
candidate at the $z$ of the wire from any other track by a minimum distance of
250 $\mu$m.
The types and values of the cuts are not exactly the same as in the muon analysis
due to the very different background conditions. 

The detached selection yields 8 events upstream of the primary interaction 
region (pure combinatorial background) and 19 downstream events
(see Fig.~\ref{fig:eedetached}).  
Among the downstream candidates, 10 events are found in the \jpsi\ mass window 
($2.8\ \mgev\ <m_{\ee}<3.3\ \mgev$). 

%
%
%
%
\begin{figure}
\resizebox{0.75\textwidth}{!}{%
\includegraphics{\Figdir 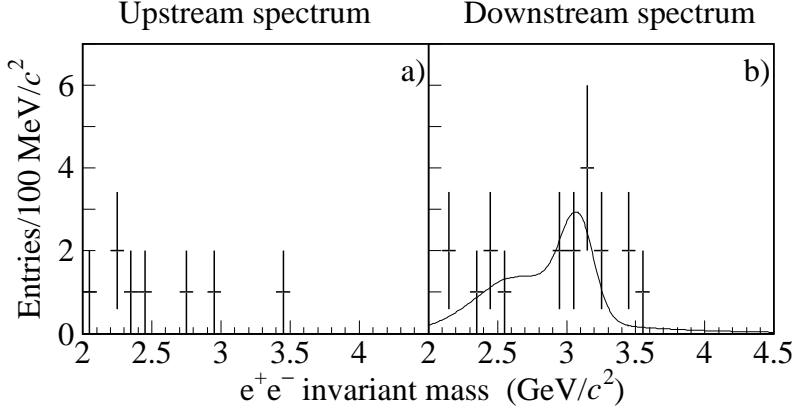}}
\caption{
The upstream (a) and downstream (b) \ee\ invariant mass spectra
after detached event selection. The downstream curve shows the result
of the unbinned likelihood fit, in which the yields of background and signal contributions, as defined in the text, were left as free parameters.
}
\myfiglabel{fig:eedetached} 
\end{figure}

Similarly to the muon analysis, an unbinned likelihood fit is performed on the
invariant mass spectrum of the detached  downstream \ee\ candidates. The shape
of the signal is taken from simulated \bjpsi\ decays, while the background
shape is a combination of the shapes obtained 
from simulated double semileptonic bottom quark decays and from
pure combinatorial (upstream) events. 
The result of the likelihood fit is shown in
Fig.~\ref{fig:eedetached}(b), yielding 8.6$^{+3.9} _{-3.2}$ \bjpsi\ events.
When the background shape used in the fit is replaced by a pure
combinatorial background shape or by a pure double semileptonic
\bbar\ background, a $\pm 7$\% variation is observed in the
number of \bjpsi\ events. This contribution is included in the systematic
error of our measurement.

The expected background from prompt \jpsi\ decays is of less than 0.2 events 
at the 90\% C.L. in the  whole downstream region. 
As in the \mm\ case, the estimated background 
yields from semileptonic charm and bottom quark decays, together with the 
expected combinatorial background level (8 events seen upstream) and the 
fitted signal, are compatible with the 19 events observed downstream 
of the primary interaction region.

Different cut optimization techniques and assumptions have been tested to
verify the stability of the signal.
The optimizations are performed simultaneously on the three detached vertex 
cuts ($\Delta z$, $I_w$ and the isolation cut) using the background from real
data and the downstream \bjpsiee\ events from MC.
Independently on the optimization criteria, a \jpsi\ signal with 
significance greater than 2 $\sigma$ 
is always observed in the downstream part of the spectrum,
while a visible \jpsi\  signal is never present in the upstream part.
In Fig.~\ref{fig:eedetached2}, the selected detached events are displayed 
in a scatter plot of the invariant mass versus the measured decay length 
($\Delta z$): a clustering is observed around the \jpsi\ invariant mass 
for large $\Delta z$ values in the region downstream of the primary 
interaction. 
 
%
%
%
%
\begin{figure}
\resizebox{0.75\textwidth}{!}{%
\includegraphics{\Figdir 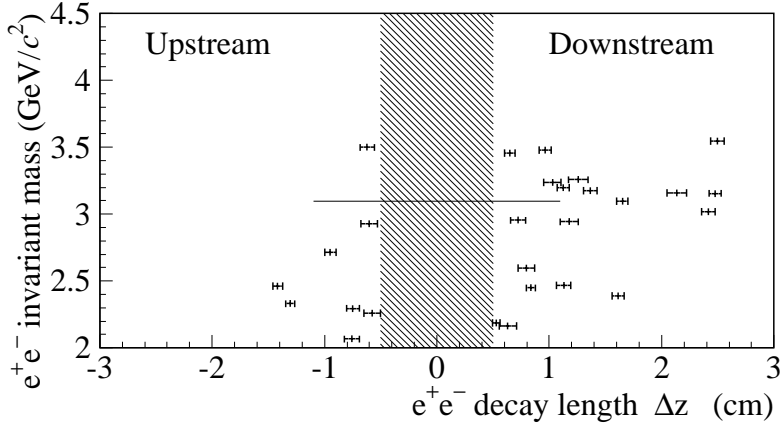}}
\caption{
The scatter plot of \ee\ invariant masses versus
the measured decay length ($\Delta z$) for the selected detached events. 
The shaded
region is removed by the $\Delta z$ cut. The horizontal line shows the mean \jpsi\ invariant mass value. A clear clustering of events around the \jpsi\ mass with large $\Delta z$ is observed in the downstream sample.
}
\myfiglabel{fig:eedetached2} 
\end{figure}

To confirm the \B\ assignment of the selected events, 
an unbinned maximum likelihood fit
is performed on the decay length measurements.
A mean decay length of 
$0.81 \pm 0.03$ cm on \bbar\ events is expected from MC. 
We measure $1.0 \pm 0.3$ cm  for the 10 downstream events in the \jpsi\ region
($2.8\ \mgev\ <m_{\ee}<3.3\ \mgev$), 
in good agreement with the \bbar\  interpretation, while the 8 upstream
background events yield a mean decay length of $0.36 \pm 0.13$ cm (measured
using $-\Delta z$). To further verify that the selected events have 
features compatible with \B\ decays, 
we performed a visual inspection of the candidates studying
extra detached vertices (from the other \B\ decay)  
and extra tracks attached to the \jpsi\ vertex.  Both
categories of events are observed, and their yields are compatible with MC
expectations. Within the limits of the available statistics, the \jpsi\ \xf\ 
 and  \pt\  distributions are also compatible with the \bjpsi\ interpretation. 

From MC simulation, we obtain the efficiency terms entering in the cross
section measurement (Eq.~\ref{eq:sigbbar}): 
$\effR \cdot \effBz = 0.44\pm 0.02$.
The corresponding \bbar\ cross section measured in the \ee\ channel is 
$\dsigbbar = \sigB/A = 38^{+18}_{-15}$
nb/nucleon, obtained by using the weighted average of our target
materials. All the parameters used in Eq.~\ref{eq:sigbbar} are summarized in
Table~\ref{tab:comp}.

%
%
%
%
%
\begin{table}[th]
\begin{tabular}{|l|cc|}
\hline 
       & \mm\    & \ee\  \\
       & channel & channel \\ \hline
\hline
Target  & \multicolumn{2}{c@{\ \ \ }|}{77\% C($A$=12) +23\% Ti($A$=48)} \\
Interaction rate & \multicolumn{2}{c@{\ \ \ }|}{5 MHz} \\
Beam energy & \multicolumn{2}{c@{\ \ \ }|}{920 GeV} \\
$\sqrt{s}$  &  \multicolumn{2}{c@{\ \ \ }|}{41.6 GeV} \\
$\alpha$ & \multicolumn{2}{c@{\ \ \ }|}{$0.955 \pm 0.005$} \\
$\sigma(\jpsi)$  & \multicolumn{2}{c@{\ \ \ }|}{$357 \pm 28$ nb/nucleon} \\
Prompt \jpsi\ ($N_P$) &  $2880 \pm 60$   & $5710 \pm 380$ \\
Detached \jpsi\ ($N_B$) & $1.9^{+2.2} _{-1.5}$ &  $8.6^{+3.9} _{-3.2}$  \\
$\effR \cdot \effBz $  & $ 0.41\pm 0.01$ & $0.44 \pm 0.02$ \\
Br$(b \bar b \to \jpsi X)$ & \multicolumn{2}{c@{\ \ \ }|}{($2.32 \pm 0.20 \ $)\%} \\[0.1cm]
\dsigbbar\   & $16^{+18} _{-12}$ nb/nucl. & $38^{+18} _{-15}$ nb/nucl.\\ \hline
Combined \dsigbbar\  & \multicolumn{2}{c@{\ \ }|}{$30^{+13} _{-11} 
\mathrm{(stat)} \pm
  6 \mathrm{(sys)}$ nb/nucleon}\\
Combined \sigbbar\  & \multicolumn{2}{c@{\ \ }|}{$32^{+14} _{-12} 
\mathrm{(stat)} \, ^{+6} _{-7} \mathrm{(sys)}$ nb/nucleon}\\
\hline 

\end{tabular}
\caption{\label{tab:comp} The parameters entering into the 
\sigbbar\ measurement (Eq.~\ref{eq:sigbbar}). 
}
\end{table}    
%
%
\section{Combined Cross Section Measurement}

The two measurements, 
$\dsigbbar=16^{+18} _{-12}$ nb/nucleon and 
$\dsigbbar = 38^{+18}_{-15}$ nb/nucleon, obtained 
in the muon and electron channels, respectively,  
are compatible within statistical uncertainties.
In order to extract the maximum information on the
\bbar\ production cross section from our data, we combine the 
\mm\ and \ee\ likelihoods in a four parameter likelihood maximization 
(\dsigbbar , \mm\ background slope, \mm\ and \ee\ background yields)
on the detached candidates. 
The fit provides our final result of 
the \bbar\ production cross section:
\begin{equation}
\dsigbbar\  = 30 ^{+13} _{-11} \mathrm{(stat)} \ \mathrm{nb/nucleon},
\end{equation}
where the quoted uncertainty has been estimated directly from the fit
(see Fig.~\ref{fig:xslikelihoods}).

%
%
%
%
\begin{figure}
\resizebox{0.75\textwidth}{!}{%
\includegraphics{\Figdir 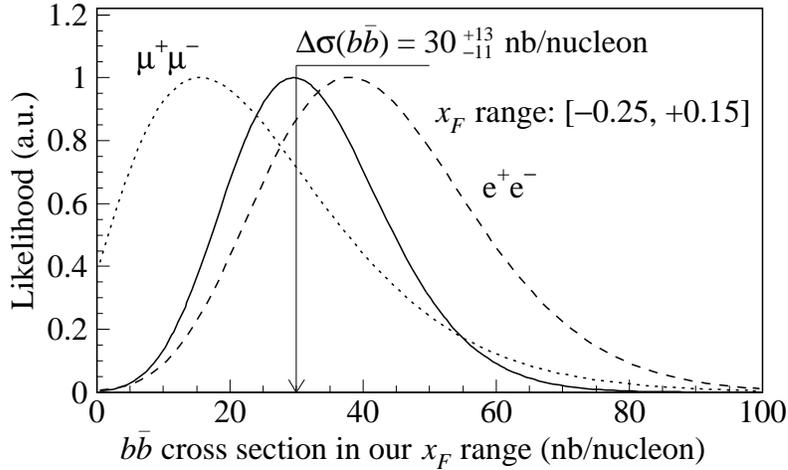}}
\caption{The likelihood fits for the \bbar\ production cross section 
in our \xf\ range (\dsigbbar) using the \mm\ and \ee\ events separately 
(dotted and dashed line respectively) and in a combined analysis (solid line).}
\myfiglabel{fig:xslikelihoods}    
\end{figure}

The main sources of systematic uncertainty in the present measurement, 
which are not related to the final \bbar\ statistics, are due to
the prompt \jpsi\ cross section reference (11\%), the branching ratio 
Br$(b \bar b \to \jpsi X)$ (9\%), the trigger and detector 
simulation (5\%), the prompt \jpsi\ MC production models (2.5\%), the 
\bbar\ MC production models (5\%), the prompt \jpsiee\ counting (5\%) and 
the carbon-titanium difference in efficiencies (1.7\%). Other contributions are below 
the 1\% level.
Uncertainties stemming from the background shapes used in the 
maximum likelihood fits on the invariant masses 
and from the cut values are dominated by the low statistics
of observed detached events. For these sources we assign conservative 
uncertainties of $^{+10} _{-24}$\% and 13\% to the \mm\ and \ee\ channels, respectively.
The overall systematic uncertainty for our measurement, averaged 
over the muon and electron channels, is of $^{+20} _{-23}$ \%.

To compare our measurement with theoretical predictions, we extrapolate the 
\dsigbbar\ measurement to the full \xf\ range, relying on the \bbar\ production
and decay model described in Sect.~\ref{sect:mc} which foresee that 92\% of
\jpsi\ from \B\ decays are produced in our \xf\ range. 
We obtain the total \bbar\ production cross section:

\begin{equation}
\sigbbar\ = 32 ^{+14} _{-12} \mathrm{(stat)} \, ^{+6} _{-7} \mathrm{(sys)} \ \mathrm{nb/nucleon}.
\end{equation}

In Fig.~\ref{fig:bbarexpth}, this result is compared with 
the latest QCD calculations~\cite{nason1998,vogt2001}
beyond next-to-leading order (NLO).
The two predicted values at 920 GeV proton beam are, respectively, 
$\sigbbar\ = 25^{+20} _{-13}$ nb/nucleon~\footnote{Value based on Ref.~\cite{nason1998}, 
updated with the parton 
distribution function in Ref.~\cite{mrst}.}
and
$\sigbbar\ = 30 \pm 13$ nb/nucleon, 
in good agreement with our measurement.
In the same figure, the E789 \cite{E789bb} and E771 \cite{E771bb}
experimental results obtained with 800 GeV proton interactions on 
Au and Si, respectively, are plotted and are seen to be compatible 
(Fig.~\ref{fig:bbarexpth}).

%
%
%
%
\begin{figure}
\resizebox{0.75\textwidth}{!}{%
\includegraphics{\Figdir 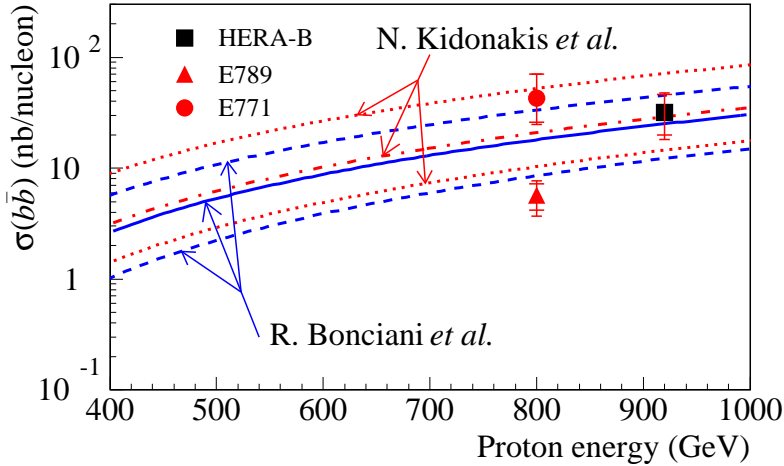}}
\caption{The comparison of the \hb (2000) \sigbbar\ value with 
other experiments and
with the theoretical predictions of R. Bonciani {\it et al.} \cite{nason1998} 
updated with the NNLL parton distribution function in \cite{mrst}
(solid line: central value, dashed lines: upper and lower bounds) and
N. Kidonakis {\it et al.} \cite{vogt2001} (dot-dashed line: central value, 
dotted lines: upper and lower bounds).}
\myfiglabel{fig:bbarexpth}    
\end{figure}

\section{Conclusions}
\label{sec:concl}

Events coming from \bjpsill\ decays have been identified in a sample of $\approx$1.35 
million dilepton triggered events, acquired in a short physics run during the 
\hb commissioning period in summer 2000. 
The data analysis results in the identification of $1.9^{+2.2}_{-1.5}$ 
\bjpsimm\ candidates and $8.6 ^{+3.9} _{-3.2}$ \bjpsiee\ candidates.

From these candidates, we compute the \bbar\ production cross section 
by normalizing to the known prompt \jpsi\ cross section. 
In the \jpsi\ kinematic range $-0.25< \xf <0.15 $,
we obtain  $\dsigbbar=16 ^{+18} _{-12}$ nb/nucleon and 
$\dsigbbar = 38^{+18}_{-15}$ nb/nucleon in the muon and electron channels,
respectively. Within statistical errors, the two results are compatible. 
The combined result of the \bbar\ production cross section measured by \hb 
at 920 GeV using $p$C and $p$Ti interactions in our \xf\ range is
$\dsigbbar\ = 30 ^{+13} _{-11} \mathrm{(stat)} \pm 6 \mathrm{(sys)} \ \mathrm{nb/nucleon}$.
Extrapolating this measurement to the full \xf\ range, we obtain the
total \bbar\ production cross section:

\begin{equation}
\sigbbar\ = 32 ^{+14} _{-12} \mathrm{(stat)} \, ^{+6} _{-7} \mathrm{(sys)} \ \mathrm{nb/nucleon}.
\end{equation}

This result is compatible with the existing measurements~\cite{E789bb,E771bb}
and in agreement with the most recent
QCD predictions~\cite{nason1998,vogt2001} beyond NLO.

\section*{Acknowledgments}
\setcounter{footnote}{2}
\renewcommand\thefootnote{\fnsymbol{footnote}}

We express our gratitude to the DESY laboratory and to the DESY
accelerator group for their strong support since the conception of the
\hb experiment. The \hb experiment would not have been possible
without the enormous effort and commitment of our technical and
administrative staff. It is not possible to list here the many
individuals who have contributed to \hbp . We are especially grateful
to the following persons and their groups: G.~Avoni, C.~Baldanza, J.~Bizzell, 
A.~Cotta-Ramusino, F.~Czempik, I.~D'Antone, J.~Davila, J.~Dicke,
A.~Donat, U.~Dretzler, A.~Epifantsev, S.~Fricke, W.~Funk, A.~Gutierrez, 
F.~Hansen, M.~Harris, S.~Hennenberger, J.~Hogenbirk, 
M.~Jablonski, V.~Kiva, M.~Kolander, Y.~Kolotaev, L.~Laptin, H.~Leich, 
H.~L\"udecke, Q.~Li, K.~Y.~Liu, P.~Liu, C.~Lu, K.~Ludwig, J.~McGill, 
E.~Michel, N.~Murthy, E.~Novikov, S.~Omeltchuk, D.~Padrazo, 
H.~B.~Peters, P.~Pietsch, M.~Pohl, N.~Ratnikova,
A.~Rausch, W.~Reinsch, P.~Rose, I.~Rostovtseva, R. Rusnyak, W.~Sands, P.~Solc,
S.~Starostin$^{\dagger}$\footnote[0]{$^\dagger$ \it deceased }, 
K.-H.~Sulanke, V.~Tchoudakov, M.~Tkatch, K.~Wagner, 
P.~Wegner, V.~Zerkin, E.~Zimmer-Nixdorf. We thank the external 
\hb referees R. Forty, D.~Froidevaux, \mbox{R.-D.~Heuer},
K.~Jakobs and J.~Jaros for many stimulating discussions and
suggestions. We are indebted to our administrative staff U.~Djuanda and
I.~Kerkhoff for their continuous assistance.

In the preparation of this paper, we have benefited from many useful
discussions with M.~Mangano, P.~Nason, and R.~Vogt on the theory of
\bbar\ production.

%

\end{document}

%% file: macros.tex
\newcommand{\NIM}[3]       {Nucl.\ Instr.\ Methods~{\bf A#1} (#2) #3}

\newcommand{\NPB}[3]       {Nucl.\ Phys.~{\bf B#1} (#2) #3}
\newcommand{\PLB}[3]       {Phys.\ Lett.~{\bf B#1} (#2) #3}

\newcommand{\PRD}[3]       {Phys.\ Rev.~{\bf D#1} (#2) #3}
\newcommand{\PRL}[3]       {Phys.\ Rev.\ Lett.~{\bf #1} (#2) #3}

\newcommand{\EPJ}[3]       {Eur.\ Phys.\ J.~{\bf C#1} (#2) #3}
\newcommand{\CPC}[3]       {Comp.\ Phys.\ Comm.~{\bf #1} (#2) #3}
\newcommand{\hb} {\mbox{\sffamily HERA \protect\rule[.5ex]{1.ex}{.11ex} B}\ }
\newcommand{\hbp}{\mbox{\sffamily HERA \protect\rule[.5ex]{1.ex}{.11ex} B}}
\newcommand{\ra} {\mbox{$\mskip 3mu \rightarrow \mskip 5mu$}}
\newcommand{\eff}{\ensuremath{\varepsilon}}

\newcommand{\effBz}{\ensuremath{\eff_B^{\Delta z}}}

\newcommand{\effR}{\ensuremath{\eff_R}}
\newcommand{\bbjX}{\ensuremath{\bbar\ra\jpsiX}}

\newcommand{\Br}[1]{\ensuremath{{\rm  Br}(#1)}}
\newcommand{\sigB}{\ensuremath{\sigma_B^A}}
\newcommand{\sigP}{\ensuremath{\sigma_P^A}}

\newcommand{\B}{\ensuremath{{b}}}

\newcommand{\bjpsi}{\ensuremath{\B \to\jpsi}}
\newcommand{\bjpsill}{\ensuremath{\B \to\jpsi\to l^+l^-}}
\newcommand{\bjpsiee}{\ensuremath{\B \to\jpsi\to e^+e^-}}
\newcommand{\bjpsimm}{\ensuremath{\B \to\jpsi\to \mu^+\mu^-}}
\newcommand{\bbar}{\ensuremath{\B\overline{\B}}}
\newcommand{\sigbbar}{\ensuremath{\sigma(\bbar)}}

\newcommand{\dsigbbar}{\ensuremath{\Delta \sigma(\bbar)}}
\newcommand{\ee}{\ensuremath{e^+e^-}}
\newcommand{\mm}{\ensuremath{\mu^+\mu^-}}
\newcommand{\jpsi}{\ensuremath{J/\psi}}
\newcommand{\jpsiX}{\ensuremath{\jpsi X}}
\newcommand{\jpsill}{\ensuremath{ \jpsi\to l^+l^-}}
\newcommand{\jpsiee}{\ensuremath{ \jpsi\to e^+e^-}}
\newcommand{\jpsimm}{\ensuremath{ \jpsi\to \mu^+\mu^-}}
\newcommand{\xf}{\ensuremath{x_{ F}}}
\newcommand{\pt}{\ensuremath{p_{ T}}}

\newcommand{\egev}{\ensuremath{\, \mathrm{GeV}}}
\newcommand{\mgev}{\ensuremath{\, \mathrm{GeV}/c^2}}

%% file: authors.tex
\normalsize
\begin{center}
I.~Abt$^{28}$, A.~Abyzov$^{26}$, M.~Adams$^{11}$, H.~Albrecht$^{13}$,
V.~Amaral$^{8}$, A.~Amorim$^{8}$, S.~J.~Aplin$^{13}$, A.~Arefiev$^{25}$,
I.~Ari\~no$^{2}$, M.~Atiya$^{36}$
% ...

\end{center}

\small
\begin{center}
\address{
$^{1}${Department of Physics, University of Texas, Austin, TX 78712-1081, USA$^{a}$} \\ 
$^{2}${Department ECM, Faculty of Physics, University of Barcelona, E-08028 Barcelona, Spain~$^{b}$} \\
$^{3}${Institute for High Energy Physics, Beijing 100039, P.R. China} \\
$^{4}${Institute of Engineering Physics, Tsinghua University, Beijing 100084, P.R. China} \\
$^{5}${Institut f\"ur Physik, Humboldt-Universit\"at zu Berlin, D-10115 Berlin, Germany~$^{c}$} \\
$^{6}${Dipartimento di Fisica dell' Universit\`{a} di Bologna and INFN Sezione
  di Bologna, I-40126 Bologna, Italy} \\
$^{7}${Department of Physics, University of Cincinnati, Cincinnati, Ohio 45221, USA$^{a}$} \\
$^{8}${LIP Coimbra and Lisboa, P-3004-516 Coimbra,  Portugal~$^{d}$} \\
$^{9}${Niels Bohr Institutet, DK 2100 Copenhagen, Denmark~$^{e}$} \\
$^{10}${Department of Physics and Astronomy, Wayne State University, Detroit, MI 48202, USA~$^{a}$} \\
$^{11}${Institut f\"ur Physik, Universit\"at Dortmund, D-44227 Dortmund, Germany~$^{c}$} \\
$^{12}${Northwestern University, Evanston, Il 60208, USA$^{a}$} \\
$^{13}${DESY, D-22603 Hamburg, Germany} \\ 
$^{14}${Institut f\"ur Experimentalphysik, Universit\"at Hamburg, D-22761 Hamburg, Germany~$^{c}$} \\
$^{15}${Max-Planck-Institut f\"ur Kernphysik, D-69117 Heidelberg, Germany~$^{c}$} \\
$^{16}${Physikalisches Institut, Universit\"at Heidelberg, D-69120 Heidelberg, Germany~$^{c}$} \\
$^{17}${Department of Physics, University of Houston, Houston, TX 77204, USA~$^{a,f}$} \\
$^{18}${Institute for Nuclear Research, Ukrainian Academy of Science, 03680 Kiev, Ukraine~$^{g}$} \\
$^{19}${J.~Stefan Institute, 1001 Ljubljana, Slovenia} \\
$^{20}${University of Ljubljana, 1001 Ljubljana, Slovenia} \\
$^{21}${University of Maribor, 2000 Maribor, Slovenia} \\
$^{22}${University of California, Los Angeles, CA 90024, USA$^{h}$} \\
$^{23}${Lund University, S-22362 Lund, Sweden} \\
$^{24}${Lehrstuhl f\"ur Informatik V, Universit\"at Mannheim, D-68131 Mannheim, Germany } \\
$^{25}${Institute of Theoretical and Experimental Physics, 117259 Moscow, Russia~$^{i}$} \\
$^{26}${Joint Institute for Nuclear Research Dubna, 141980 Dubna, Moscow region, Russia} \\
$^{27}${Moscow Physical Engineering Institute, 115409 Moscow, Russia} \\
$^{28}${Max-Planck-Institut f\"ur Physik, Werner-Heisenberg-Institut, 
D-80805 M\"unchen, Germany~$^{c}$} \\
$^{29}${Dept. of Physics, University of Oslo, N-0316 Oslo, Norway $^{j}$} \\
$^{30}${Fachbereich Physik, Universit\"at Rostock, D-18051 Rostock, Germany~$^{c}$} \\
$^{31}${Fachbereich Physik, Universit\"at Siegen, D-057068 Siegen, Germany~$^{c}$} \\
$^{32}${Universiteit Utrecht/NIKHEF, 3584 CB Utrecht, The Netherlands~$^{k}$} \\
$^{33}${NIKHEF, 1009 DB Amsterdam, The Netherlands~$^{k}$} \\
$^{34}${DESY Zeuthen, D-15738 Zeuthen, Germany} \\
$^{35}${Physik-Institut, Universit\"at Z\"urich, CH-8057 Z\"urich, Switzerland~$^{l}$} \\
$^{36}${Brookhaven National Laboratory, Upton, NY 11973, USA} \\
$^{37}${visitor from Dipartimento di Energetica dell' Universit\`{a} di Firenze and INFN Sezione di Bologna, Italy} \\
$^{38}${visitor from P.N.~Lebedev Physical Institute, 117924 Moscow B-333, Russia} \\
$^{39}${visitor from Moscow Physical Engineering Institute, 115409 Moscow, Russia} \\
$^{40}${visitor from Institute of Nuclear Power Engineering, 249030, Obninsk, Russia} \\
$^{41}${visitor from Institute for High Energy Physics, Protvino, Russia} \\
$^{42}${visitor from Institute for Nuclear Research, INRNE-BAS, Sofia, Bulgaria} \\
$^{43}${visitor from High Energy Physics Institute, 380086 Tbilisi, Georgia} \\
$^{44}${visitor from Yerevan Physics Institute, Yerevan, Armenia} 
}

\vspace{5mm}
$^{a}${\it supported by the U.S. Department of Energy (DOE)} \\ 
$^{b}${\it supported by the CICYT contract AEN99-0483} \\   
$^{c}${\it supported by the Bundesministerium f\"ur Bildung und Forschung, FRG, under contract numbers 05-7BU35I, 05-7DO55P, 05 HB1HRA, 05 HB1KHA, 05 HB1PEA, 05 HB1PSA, 05 HB1VHA, 05 HB9HRA, 05 7HD15I, 05 7HH25I,  05 7MP25I, 05 7SI75I } \\
$^{d}${\it supported by the Portuguese Funda\c{c}\~ao para a Ci\^encia e Tecnologia} \\
$^{e}${\it supported by the Danish Natural Science Research Council} \\ 
$^{f}${\it supported by the Texas Advanced Research Program} \\ 
$^{g}${\it supported by the National Academy of Science and the Ministry of Education and Science of Ukraine} \\
$^{h}${\it supported by the U.S. National Science Foundation Grant PHY-9986703} \\ 
$^{i}${\it supported by the Russion Fundamental Research Foundation under grant RFFI-00-15-96584 and the BMBF via the Max Planck Research Award} \\ 
$^{j}${\it supported by the Norwegian Research Council} \\
$^{k}${\it supported by the Foundation for Fundamental Research on Matter (FOM), 3502 GA Utrecht, The Netherlands} \\ 
$^{l}${\it supported by the Swiss National Science Foundation} 

\end{center}

\normalsize